\documentclass[a4paper,twoside]{article}

\usepackage{csquotes}
\usepackage{epsfig}
\usepackage{subcaption}
\usepackage{calc}
\usepackage{amssymb}
\usepackage{amstext}
\usepackage{amsmath}
\usepackage{amsthm}
\usepackage{multicol}
\usepackage{pslatex}
\usepackage{apalike}
\usepackage{algorithm2e}
\usepackage[bottom]{footmisc}
\usepackage{multirow}
\usepackage{mathtools}
\usepackage{subcaption}
\usepackage{SCITEPRESS}

\begin{document}
\title{Action Duration Generalization for Exact Multi-Agent Collective Construction}
\author{
\authorname{Martin Rameš\sup{1}\orcidAuthor{0009-0000-3301-6269} and Pavel Surynek\sup{1}\orcidAuthor{0000-0001-7200-0542}}
\affiliation{\sup{1}Faculty of Information Technology, Czech Technical University, Thákurova 9, 160 00 Prague 6, Czech Republic}
\email{\{ramesmar, pavel.surynek\}@fit.cvut.cz}
}


\abstract{This paper addresses exact approaches to multi-agent collective construction problem which tasks a group of cooperative agents to build a given structure in a blocksworld under the gravity constraint. We propose a generalization of the existing exact model based on mixed integer linear programming by accommodating varying agent action durations. We refer to the model as a fraction-time model. The generalization by introducing action duration enables one to create a more realistic model for various domains. It provides a significant reduction of plan execution duration at the cost of increased computational time, which rises steeply the closer the model gets to the exact real-world action duration. We also propose a makespan estimation function for the fraction-time model. This can be used to estimate the construction time reduction size for the purpose of cost-benefit analysis. The fraction-time model and the makespan estimation function have been evaluated in a series of experiments using a set of benchmark structures. The results show a significant reduction of plan execution duration for non-constant duration actions due to decreasing synchronization overhead at the end of each action. According to the results, the makespan estimation function provides a reasonably accurate estimate of the makespan.}

\onecolumn \maketitle \normalsize \setcounter{footnote}{0} \vfill

\section{\uppercase{Introduction}}
The multi-agent collective construction (MACC) problem tasks a group of cooperative agents to build a given structure in a blocksworld. Agents can pick up, move, and place blocks, which are used as the only building material for a three-dimensional structure. Both blocks and agents are moving under the condition of gravity. The problem aims to determine a collision-free plan for the agent movement, which would perform the construction task while minimizing the execution time (makespan) and the sum of durations the agents spend on the grid (sum-of-costs).

\begin{figure}
    \centering
    \begin{subfigure}[b]{\columnwidth}
        \centering
        \includegraphics[scale=0.25]{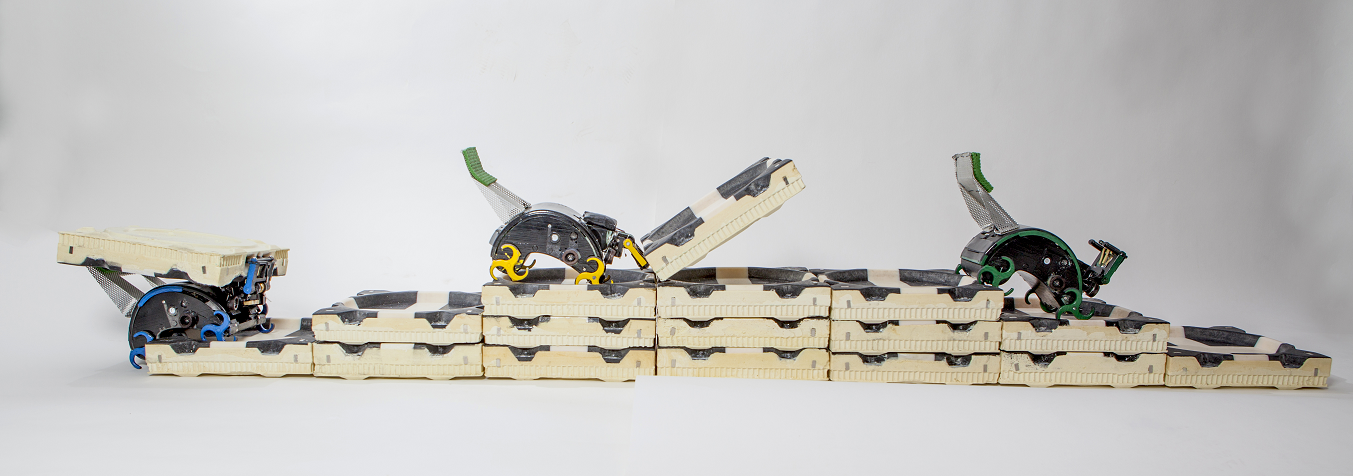}
        \caption{"Termes robot 01" by Forgemind ArchiMedia is licensed under CC BY 2.0 \cite{ImageTERMES}.}
    \end{subfigure}
    \begin{subfigure}[b]{\columnwidth}
        \centering
        \includegraphics[scale=1]{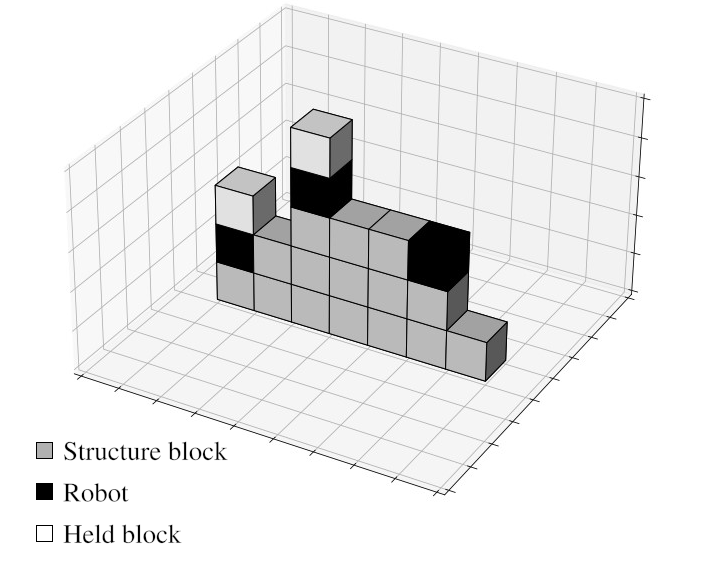}
        \caption{Visualization of the world state in the MACC problem.}
    \end{subfigure}
    \caption{Example of the TERMES system and its MACC representation. Three robotic agents are building a long stair structure. The middle agent is performing a deliver-block action.}
    \label{fig:TERMES2MACC}
\end{figure}

Previously, such problems were solved using only heuristic approaches with no proof of optimality. Recently, a new branch of research emerged, aiming to study optimal solutions for the MACC problem. This paper aims to further this research by providing a generalization of the currently best optimal model -- MILP model by \cite{koenigexact} -- further referred to as the constant-time model. The generalization aims to allow agent actions to differ in duration and to better map real-world multi-agent systems.

The constant-time model \cite{koenigexact} exactly optimizes the multi-agent construction plan to achieve the MACC problem solution with optimal makespan and sum-of-costs as primary and secondary optimization criteria, respectively. The model, however, assumes all actions to have the same unit duration. In real life, this is not the case and the robots are forced to wait for the end of the longest action. This includes the TERMES robots \cite{petersen2011termes}, which serve as inspiration for the constant-time model. Notably, the relative difference between average block pick-up time (15s) and placement time (24s), measured in \cite{petersen2011termes}, is 46\%. The paper also shows that agents carrying a block move slower, as is the case when moving on more complex unstructured terrain, such as gravel or grass.

We propose a generalization of the constant-time model to better utilize the information about mean action durations. The generalized model, further referred to as the fraction-time model, will accept structure description along with action duration assignment as input and provide a MACC plan as output. The structure description will be passed as a height map, the same as in the constant-time model. The action duration assignment will assign positive rational numbers as action durations. These rational numbers will be scaled to integers by multiplying them by the least common multiple of their denominators as the only preprocessing step, keeping the generalized MILP model simple. We modify the constant-time model to include the action durations. The modification is non-trivial, as the generalized model needs to prevent mid-action agent collisions and maintain the network flow substructure of the constant-time model to remain computationally viable.

We demonstrate the improvement of expected plan execution duration on three different instances with three non-constant action durations, including times derived from measurement on the TERMES robots in \cite{petersen2011termes}. We provide two lower bounds for makespan estimation, one based on fraction-time model relaxation and one based on less detailed action duration assignment for the fraction-time model. We also show that the less-detailed action duration assignment can be utilized for makespan upper bound estimation. Finally, we show the potential to compute plans with a reduced temporary block placement count, which has previously been done heuristically by \cite{TACR}. To solve the fraction-time MILP model, we use the state-of-the-art Gurobi solver \cite{Gurobi}.

\section{\uppercase{Related Work}}
The MACC problem is part of a wider research branch, which studies various forms of collaborative construction. These forms include -- for instance -- self-assembly, where the agents themselves act as building blocks, once they reach their intended position \cite{SphereRobots,CubeRobots}. Another approach utilizes quadrotors to carry beams and columns, which are used to assemble the structure \cite{Quadrotors}. Robotic arms with connectors on both sides have been proposed to build and move atop lightweight lattice structures like an inchworm \cite{NASAinchworm}. Similarly, a layer-by-layer multi-agent construction using two-ended robotic arms and intelligent blocks has been proposed by \cite{studentInchworm}.

All these approaches focus on robots with unique capabilities and limitations, which require dedicated algorithms. We concentrate on the TERMES system, a termite-inspired multi-agent system by \cite{petersen2011termes}. The TERMES robots can pick up, carry, and place similar-sized blocks. Each robot can carry at most one block at a time, and climb the difference of one block. The robots are capable of building complex structures based on high-level instructions and simple sensor feedback. These instructions are -- move forward by one block, turn 90° left, turn 90° right, pick up the block in front of the robot, and deliver a carried block at a position in front of the robot \cite{petersen2011termes}. The MACC problem discretizes the movement of the TERMES robot and makes the problem partially hardware-independent by limiting the robots to the high-level actions described above. Image \ref{fig:TERMES2MACC} shows the TERMES robots and the MACC discretization of the structure.

Currently, there are only two approaches that can solve the MACC problem makespan-optimally \cite{koenigexact} -- a mixed integer linear programming (MILP) model and a constraint programming (CP) model. The results of both models in the paper match on finished instances, with the MILP model proving significantly lower computation times. Both models assume a constant action duration.

There are also multiple heuristic/sub-optimal approaches to the MACC problem. The Compiler for Scalable Construction by the TERMES Robot Collective provides solutions heuristically minimizing the number of agents, who pass through the construction site \cite{scalableConstructionCompiler}. Another approach presents an algorithm minimizing the number of pick-up and deliver actions, performing dynamic programming on a spanning tree to find paths for a TERMES robot \cite{multiagentTACR}.

There are also two approaches, which utilize a solver/planner but sacrifice the optimal solution to gain order-of-magnitude better computation speed. The first decomposes the target structure into substructures, exactly optimizing the construction of substructures using a MILP model \cite{srinivasan2023multiagent}. The second model uses a hierarchical planning procedure \cite{pddl}. The model first uses propositional logic and graph search to identify a sequence of block pick-up and deliver actions building the target structure. The algorithm then builds an action dependency graph for the sequence actions, assigns the actions to the agents, and plans collision-free paths using a multi-agent path-finding planner.

None of the above models guarantee an optimal solution, though the last two utilize the optimal solution of another model for comparison of their results.

\section{\uppercase{The Proposed Generalized MILP Model}}
To allow more precise modeling of non-constant action durations of real-world robots, we propose the fraction-time model. The proposed model replaces the timestep $t$ of each action by $t_s$ and $t_e$, which denote the timestep of action start (inclusive) and action end (exclusive), respectively. This notation mirrors non-preemptive scheduling notation, where the action is executed within a time interval $[{t_s,t_e})$.

The proposed model uses the non-standard notation of \cite{koenigexact}. For set $\mathcal{U}$ of tuples $(u_1, u_2, \dots , u_n)$ with length $n \in \mathbb{Z}_+$ notation $\mathcal{U}_{u_1, u_2, \dots , u_n}$ is shorthand for $\{(u_1', u_2', \dots , u_n'): u_1' = u_1 \allowbreak\wedge u_2' = u_2 \allowbreak\wedge \dots \wedge u_n' = u_n\}$. Let $*$ be a wildcard symbol matching any value at position in the tuple it is used in (meaning $\mathcal{U}_{*, u_2, \dots , u_n}$ is shorthand for $\{(u_1', u_2', \dots , u_n'): u_2' = u_2 \allowbreak\wedge \dots \wedge u_n' = u_n\}$ and $\mathcal{U}_{u_1, u_2, *, \dots , *}$ is shorthand for $\{(u_1', u_2', \dots , u_n'): u_1' = u_1 \allowbreak\wedge u_2' = u_2\}$).

Let $\widehat{a}$ be shorthand for $\{0, \dots, a-1\}, a \in \mathbb{Z}_+$. Let $(X, Y)$ be the size of the building area and $(Z-1)$ be the height of the target structure in blocks, the extra layer allows travel on top of the structure. Let $\mathcal{C} = \widehat{X} \times \widehat{Y} \times \widehat{Z}$ be the set of all positions within the grid, $\mathcal{P} = \widehat{X} \times \widehat{Y}$ is the projection of $\mathcal{C}$ into the first two dimensions, $\mathcal{B} = \{(x, 0, 0): x \in \widehat{X}\} \cup \{(x, Y-1, 0): x \in \widehat{X}\} \cup \{(0, y, 0): y \in \widehat{Y}\} \cup \{(X-1, y, 0): y \in \widehat{Y}\}$ is the set of border cells at the perimeter of the building area. Let $\overline{z}_{(x, y)}$ be the target height of the block column at position $(x, y)$. Let $\mathcal{C'} = \mathcal{C} \cup \{(\text{S}, \text{S}, \text{S}), (\text{E}, \text{E}, \text{E})\}$ be a set of all agent-accessible positions -- including two special positions $(\text{S}, \text{S}, \text{S}), (\text{E}, \text{E}, \text{E})$, which symbolize the start and the end position outside the grid. Let $\mathcal{K} = \{\text{M}, \text{P}, \text{D}\}$ be a set of agent action type distinguishers -- M for move action (used for \enquote{entry}, \enquote{leave}, \enquote{move\_block}, \enquote{move\_empty} and \enquote{wait} action types), P for \enquote{pick\_up} action type and D for \enquote{deliver} action type. Let $\mathcal{N}_{(x, y)} = \{(x - 1, y), (x + 1, y), (x, y - 1), (x, y + 1)\} \cap \mathcal{P}$ be the set of neighbor positions of $(x, y)$ and $\mathcal{T} = \widehat{T}$ be the planning horizon of $T$ timesteps. The actions are tuples $i = (t_s, t_e, x, y, z, c, k, x', y', z')$, which consist of values:
\begin{itemize}
    \item start time $t_s \in \mathbb{Z}_+$ (inclusive)
    \item end time $t_e = t_s + d, d \in \mathbb{Z}_+$ (exclusive), where $d$ is the action duration
    \item start position $(x, y, z) \in \mathcal{C} \cup \{(\text{S}, \text{S}, \text{S})\}$
    \item indicator if agent carries a block at the start of the action $c \in \{0, 1\}$
    \item action type distinguisher $k \in \mathcal{K}$
    \item end position $(x', y', z') \in \mathcal{C} \cup \{(\text{E}, \text{E}, \text{E})\}$
    \begin{itemize}
        \item position of affected block for pick\_up and deliver actions (agent stays at the same position)
        \item marks agent position at the end of the action for other action types
    \end{itemize}
\end{itemize}
Let duration $d_i$ of action $i = (t_s, t_e, \dots)$ be $d_i = t_e - t_s$. Let $f_d: \mathbb{Z}_+ \times \mathcal{C'} \times \{0, 1\} \times \mathcal{K} \times \mathcal{C'} \longrightarrow \mathbb{Z}_+$ be a function of action duration, based on the rest of action tuple -- that is $t_s$, $(x, y, z)$, $c$, $k$ and $(x', y', z')$ respectively.

Let $\mathcal{A} = \{\text{entry},
\text{leave},
\text{move\_block},
\text{move\_empty},\allowbreak
\text{deliver},
\text{pick\_up},
\text{wait}\}$ be the set of all action types.

\begin{itemize}
    \item $\mathcal{Q}_\text{entry} = \{(t_s, \text{S}, \text{S}, \text{S}, c, \text{M}, x, y, z): t_s \in \widehat{T - 3} \wedge c \in \{0, 1\} \wedge (x, y, z) \in \mathcal{B}\}$
    \item $\mathcal{Q}_\text{move\_empty} = \{(t_s, x, y, z, 0, \text{M}, x', y', z'): t_s \in \{1, \dots , T - 3\} \wedge (x, y, z) \in \mathcal{C} \wedge (x', y') \in \mathcal{N}_{(x, y)} \wedge |z' - z| \leq 1\}$
    \item $\mathcal{Q}_\text{move\_block} = \{(t_s, x, y, z, 1, \text{M}, x', y', z'): t_s \in \{1, \dots , T - 3\} \wedge (x, y, z) \in \mathcal{C} \wedge (x', y') \in \mathcal{N}_{(x, y)} \wedge |z' - z| \leq 1\}$
    \item $\mathcal{Q}_\text{wait} = \{(t_s, x, y, z, c, \text{M}, x, y, z): t_s \in \{1, \dots , T - 3\} \wedge c \in \{0, 1\} \wedge (x, y, z) \in \mathcal{C}\}$
    \item $\mathcal{Q}_\text{leave} = \{(t_s, x, y, z, c, \text{M}, \text{E}, \text{E}, \text{E}): t_s \in \{2, \dots , T - 2\} \wedge c \in \{0, 1\} \wedge (x, y, z) \in \mathcal{B}\}$
    \item $\mathcal{Q}_\text{pick\_up} = \{(t_s, x, y, z, 0, \text{P}, x', y', z): t_s \in \{1, \dots , T - 3\} \wedge (x, y) \in \mathcal{P} \wedge (x', y') \in \mathcal{N}_{(x, y)} \wedge z \in \widehat{Z-1}\}$
    \item $\mathcal{Q}_\text{deliver} = \{(t_s, x, y, z, 1, \text{D}, x', y', z): t_s \in \{1, \dots , T - 3\} \wedge (x, y) \in \mathcal{P} \wedge (x', y') \in \mathcal{N}_{(x, y)} \wedge z \in \widehat{Z-1}\}$
    \item $\mathcal{Q} = \bigcup_{a \in \mathcal{A}} \mathcal{Q}_a$
\end{itemize}

Let $f_q: \mathbb{Z}_+ \times \mathcal{C'} \times \{0, 1\} \times \mathcal{K} \times \mathcal{C'} \longrightarrow \mathbb{Q}_+$ be a function, which assigns each action a duration $f_q(t_s, \dots, z'), (t_s, \dots, z') \in \mathcal{Q}$. If not specified otherwise, the $f_q$ function must be specified as part of model input. Let $m$ be the least common multiple of denominators in $\{f_q(t_s, \dots, z'): \forall (t_s, \dots, z') \in \mathcal{Q}\}$. Then, unless otherwise specified, the $f_d$ function is defined as $f_d(t_s, \dots, z') = m f_q(t_s, \dots, z'), \forall (t_s, \dots, z') \in \mathcal{Q}$.

Let us introduce the following notation to make the following equations clearer. Let $f_d(v) = f_d(t_s, x, y, z, c, k, x', y', z'), \forall v = (t_s, x, y, z, c, k, x', y', z') \in \mathcal{Q}$.

Let us define a set of all actions for each action type $a \in \mathcal{A}$ as $\mathcal{R}_a = \{(t_s, t_e, x, y, z, c, k, x', y', z'): v = (t_s, x, y, z, c, k, x', y', z') \in \mathcal{Q}_a \wedge  t_e = t_s + f_d(v)\}$.

The proposed model divides the set of agent actions into seven disjoint sets of agent action types, which in union give the new set of agent actions. Agent action types are derived from six different action subsets in the original model:

Set $\mathcal{R}_\text{entry}$ of \enquote{entry} type actions features all actions, where the agent enters from the starting position $(\text{S}, \text{S}, \text{S})$ outside the grid to a border cell. The agent might carry a block, when it enters. This action type is the only source of new blocks for the construction site.

Set of agent actions, where the agent moves to the neighbor grid position is divided into two action types -- \enquote{move\_block} for moving to the neighbor position while carrying a block and \enquote{move\_empty} for moving to the neighbor position while not carrying a block. This distinction is made for cases where the carried block requires the agent to move slower. Set $\mathcal{R}_\text{move\_empty}$ is for \enquote{move\_empty} action type and set $\mathcal{R}_\text{move\_block}$ is for \enquote{move\_block} action type. Both action types can be implemented on TERMES robots as a combination of turn and move forward actions.

Action type \enquote{wait} with action set $\mathcal{R}_\text{wait}$ has duration of $T_\text{wait} = 1$ timestep. This fixed duration is chosen, because the minimum waiting duration of agents is generally not limited.

Action type \enquote{leave} is for agents leaving the grid from a border cell (going to the end position $(\text{E}, \text{E}, \text{E})$). The associated agent action set is $\mathcal{R}_\text{leave}$. The agents can carry a block, when leaving the grid. This is the only way to remove ramp blocks from the construction site.

Set $\mathcal{R}_\text{pick\_up}$ of \enquote{pick\_up} type actions features all actions, where the agent picks up a block from a neighbor position.

Set of \enquote{deliver} type actions $\mathcal{R}_\text{deliver}$ features all actions, where the agent delivers a block to a neighbor position.

Let $a \in \mathcal{A}$ be an action type. Let $\mathcal{R}_a$ denote a set of action tuples for action type $a$. Let the set of all actions be $\mathcal{R} = \bigcup_{a \in \mathcal{A}}{\mathcal{R}_a}$. The sets $\mathcal{R}_\text{entry}, \dots, \mathcal{R}_\text{wait}$ are disjoint -- each set contains tuples with either unique position (start position $(\text{S}, \text{S}, \text{S})$ for $\mathcal{R}_\text{entry}$ and end position $(\text{E}, \text{E}, \text{E})$ for $\mathcal{R}_\text{leave}$), action type distinguisher (D for $\mathcal{R}_\text{deliver}$ and P for $\mathcal{R}_\text{pick\_up}$), or unique end position, relative to the start position (no relative movement for $\mathcal{R}_\text{wait}$ and moving to the neighbor cell with/without block ($c=1$/$c=0$) for $\mathcal{R}_\text{move\_block}$/$\mathcal{R}_\text{move\_empty}$ respectively).

Let $\mathcal{H}$ be a set of block-actions. Block-action is a tuple $(t, x, y, z, z') \in \mathcal{H}$, where $(x, y, z) \in \mathcal{C} \wedge z' \in \{z-1, z, z+1\} \cap \widehat{Z} \wedge t \in \widehat{T}$. All block-actions last one timestep, which means that only action start time $t$ is necessary. Indicators $r_i \in \{0, 1\}, i \in \mathcal{R}$ and $h_i \in \{0, 1\}, i \in \mathcal{H}$ decide, which action is part of the plan (the indicator is 1) and which action is not (the indicator is 0). For instance, $r_j = 1,\allowbreak j = (5, 7, 1, 2, 0, 1, D, 2, 2, 0) \in \mathcal{R}$ in the plan solution indicates that an agent at timestep 5 delivers block to position $(2, 2, 0) \in \mathcal{C}$ while standing at position $(1, 2, 0) \in \mathcal{C}$. The action is finished by timestep $7$.

The objective function \ref{eqn:model01} minimizes sum-of-costs, which in this context means the sum of timesteps each robot spends on the grid. The proposed model counts the \enquote{entry}-type action timesteps into the objective function, because the agent is considered to partially be on the grid, when the action starts (blocking the border cell as part of the action exclusion zone).

The exclusion zones are proposed as a measure to avoid agent collisions. Each exclusion zone is created at the start of an action, removed at the end of the same action and grants the agent performing the action exclusive access to start- and end-position columns, if the position is within grid. The exclusive access does not allow the remaining agents to perform actions, which start or end within the exclusion zone.

\begin{equation}
    \min (\sum_{i \in \mathcal{R}}{r_i d_i})
    \label{eqn:model01}
\end{equation}

\begin{equation} h_{t, x, y, z, z} = 1, \forall t \in \widehat{T}, (x, y, z) \in \mathcal{B}\label{eqn:model02}  \end{equation}
\begin{equation} h_{0, x, y, 0, 0} = 1, \forall (x, y) \in \mathcal{P}\label{eqn:model03}  \end{equation}
\begin{equation} h_{T-1, x, y, \overline{z}_{(x, y)}, \overline{z}_{(x, y)}} = 1, \forall (x, y) \in \mathcal{P} \label{eqn:model04} \end{equation}
\begin{multline}  \sum_{\mathclap{i \in \mathcal{H}_{t, x, y, *, z}}}{h_i} = \sum_{\mathclap{i \in \mathcal{H}_{t+1, x, y, z, *}}}{h_i},\quad\forall t \in \widehat{T-1}, (x, y, z) \in \mathcal{C} \label{eqn:model05}  \end{multline} 
\begin{equation}  \sum_{i \in \mathcal{H}_{t, x, y, *, *}}{h_i} = 1, \forall t \in \widehat{T}, (x, y) \in \mathcal{P} \label{eqn:model06}  \end{equation}
Constraints \ref{eqn:model02} -- \ref{eqn:model06} are the same as in the original MILP model and keep height information of construction area for every timestep. Constraint \ref{eqn:model02} forbids placement of blocks at border cells, constraint \ref{eqn:model03} starts the world devoid of blocks, constraint \ref{eqn:model04} ensures that user defined structure is finished at the end of construction, constraint \ref{eqn:model05} flows the column height from one timestep to the next (height of the block column at the end of the timestep must be equal to the height at which the column starts in the next timestep) and constraint \ref{eqn:model06} forces every position to have one height.

\begin{multline}
\sum_{i \in \mathcal{R}_{*, t, *, *, *, 0, \text{M}, x, y, z}}{r_i} +
\sum_{i \in \mathcal{R}_{*, t, x, y, z, 1, \text{D}, *, *, *}}{r_i} \\=
\sum_{i \in \mathcal{R}_{t, *, x, y, z, 0, \text{M}, *, *, *}}{r_i} +
\sum_{i \in \mathcal{R}_{t, *, x, y, z, 0, \text{P}, *, *, *}}{r_i},\\
\forall t \in \widehat{T}, (x, y, z) \in \mathcal{C}
\label{eqn:model07} \end{multline}
\begin{multline}
\sum_{i \in \mathcal{R}_{*, t, *, *, *, 1, \text{M}, x, y, z}}{r_i} +
\sum_{i \in \mathcal{R}_{*, t, x, y, z, 0, \text{P}, *, *, *}}{r_i} \\=
\sum_{i \in \mathcal{R}_{t, *, x, y, z, 1, \text{M}, *, *, *}}{r_i} +
\sum_{i \in \mathcal{R}_{t, *, x, y, z, 1, \text{D}, *, *, *}}{r_i},\\
\forall t \in \widehat{T}, (x, y, z) \in \mathcal{C}
\label{eqn:model08} \end{multline}

Constraints \ref{eqn:model07} and \ref{eqn:model08} flow the agents from one action to the next. Semi-closed interval of action execution ($[{t_s,t_e})$) is exploited for seamless transition between actions. Similarly to the base model, constraint \ref{eqn:model07} flows agents without block and ensures that the number of agents ending their action without block at position $(x, y, z)$ at timestep $t$ is the same as the number of agents without block starting their action at the same position and in the same timestep. Constraint \ref{eqn:model08} does the equivalent for agents carrying a block.

\begin{multline}
\sum_{i \in \mathcal{R}_{t_s, t_e, x, y, *, *, *, *, *, *}: t_s \le t < t_e}{r_i} +
\sum_{i \in \mathcal{R}_{t_s, t_e, *, *, *, *, *, x, y, *}: t_s \le t < t_e}{r_i} \\-
\sum_{i \in \mathcal{R}_{t_s, t_e, x, y, *, *, *, x, y, *}: t_s \le t < t_e}{r_i} \le 1,\\
\forall t \in \widehat{T}, (x, y) \in \mathcal{P}
\label{eqn:model09} \end{multline}

Constraint \ref{eqn:model09} addresses vertex collision of agents during action execution. Since the only requirement for the agent is to perform the action $a$ between $t_s$ and $t_e$, exact position of the agent is unknown and both start and end position are made into exclusion zone, where no other action can take place (for timesteps within interval $[{t_s,t_e})$). To avoid agents removing blocks under moving agents, the constraint makes the whole block column at position $(x, y)$ of both start and end of the action an exclusion zone. Separate constraint to prevent edge collisions in the base model is no longer necessary, as two actions can no longer share vertices while executing simultaneously, due to the exclusion zones.

\begin{equation}  \sum_{i \in \mathcal{R}_{t_s, t_e, *, *, *, *, *, *, *, *}: t_s \le t < t_e}{r_i} \le A,
\forall t \in \mathcal{T}\label{eqn:model11}  \end{equation}

Constraint \ref{eqn:model11} limits the number of agents on the grid. The maximum number of agents on the grid is marked as $A$ and is provided as part of model input.

\begin{multline}  \sum_{i \in \mathcal{H}_{t, x, y, z, *}}{h_i} \geq
\sum_{i \in \mathcal{R}_{t_s, t_e, x, y, z, *, *, *, *, *}: t_s \le t < t_e}{r_i},\\
\forall t \in \widehat{T}, (x, y, z) \in \mathcal{C}\label{eqn:model12}  \end{multline} 
\begin{multline} h_{t, x, y, z+1, z} = \sum_{i \in \mathcal{R}_{*, t+1, *, *, *, 0, \text{P}, x, y, z}}{r_i},\\
\forall t \in \widehat{T-1}, (x, y) \in \mathcal{P}, z \in \widehat{Z-1} \label{eqn:model13} \end{multline} 
\begin{multline}  h_{t, x, y, z, z+1} = \sum_{i \in \mathcal{R}_{*, t+1, *, *, *, 1, \text{D}, x, y, z}}{r_i},\\
\forall t \in \widehat{T-1}, (x, y) \in \mathcal{P}, z \in \widehat{Z-1} \label{eqn:model14} \end{multline}
\begin{equation}  h_i \in \{0, 1\}, \forall i \in \mathcal{H}\label{eqn:model15}  \end{equation} 
\begin{equation}  r_i \in \{0, 1\}, \forall i \in \mathcal{R}\label{eqn:model16}  \end{equation}

Constraint \ref{eqn:model12} forces agents to always stand on the highest block in the column when performing their actions. Constraints \ref{eqn:model13} and \ref{eqn:model14} govern decreases and increases in block column height, respectively. Constraint \ref{eqn:model13} ties every decrease by one block to pick\_up action. Constraint \ref{eqn:model14} does the equivalent for increase by one block and deliver action. Constraints \ref{eqn:model15} and \ref{eqn:model16} specify the variable domains.

The model is used to exactly optimize makespan and sum-of-costs, the primary and secondary optimization criterion, respectively. Optimization of makespan is done by starting at minimum possible value (estimated by a lower bound function described below) and sequentially increasing the makespan by one timestep, until a solution is found.

We expect that the most often used action duration function $f_d$ implementation will assign each action-type a fixed duration. This means, that the \enquote{deliver} action type will be assigned the duration $T_\text{deliver}$, which will apply to all actions created from $\mathcal{Q}_\text{deliver}$, the \enquote{pick\_up} action type will be assigned the duration $T_\text{pick\_up}$ and so forth -- see equation \ref{eqn:fd}. Namely, in case of the TERMES robots, the first two action type durations would be $T_\text{deliver} = 3$ and $T_\text{pick\_up} = 2$. The times assume one timestep is 10 seconds and are taken from the TERMES paper \cite{petersen2011termes}, where the block pick up time is measured to be $15 \pm 5$ s and the block delivery time (the duration of lowering and attaching a block) is measured to be $24 \pm 5$ s \cite{petersen2011termes}. The rest of the action type durations are derived in a similar fashion. The full list of TERMES action type durations is available in table \ref{tab:experimental-action-durations} in column \enquote{TERMES}.

\begin{multline}
    f_d(v) =
     \begin{cases}
       T_a &\quad\text{if } v \in \mathcal{Q}_a, a \in \mathcal{A}\\
       1 &\quad\text{otherwise.} \\ 
     \end{cases}
     \label{eqn:fd}
\end{multline}

\section{\uppercase{Makespan Estimation, Upper and Lower Bound}}
Due to the generalization, the fraction-time model is expected to be more computationally demanding than the base model, when the action durations are not constant ($\exists d_i > 1, i \in \mathcal{R}$). To allow for an informed decision, whether the use of fraction-time model is required, lower bound, heuristic estimate and upper bound of the fraction-time model makespan are presented.

A lower bound is computed as solution to the relaxed problem MACC$_{r}$. MACC$_{r}$ relaxes the requirement for the maximum number of agents (allowing unlimited number of agents), the constraint of agent collisions (allowing multiple agents to stand on one block), exclusion zone constraint (allowing agents to place blocks under other agents) and constraints limiting changes of agent vertical position (allowing agents to freely move between heights, stay on top of a column, while it is being built and place blocks to the neighbor column at the same time). The only optimization criterion for MACC$_r$ is makespan. MACC$_r$ has a trivial solution -- for each column in the building area add the number of agents equal to the column height and assign them to the column, starting at border position closest to the column. For all agents -- enter at the assigned starting position with a block, move to the assigned column (simultaneously) and one by one place the held block at the top of the assigned column. After all agents assigned to a column placed their blocks, move all those agents to the starting border cell and leave the building area. The makespan of MACC$_r$ is the lower bound of fraction-time MACC (because MACC$_r$ is the relaxation of MACC). Let $T_r$ be the optimum makespan of the problem relaxation MACC$_r$. Let $s_{(x, y)}$ be the minimum $L_1$ distance from border cell to $\mathcal{N}_{(x, y)}$ (the neighbor of $(x, y)$). Let $d_a^\text{min} = \min_{i \in \mathcal{R}_a}{d_i}, a \in \mathcal{A}$ be the minimum duration of action type $a$. Let $d_\text{move}^\text{min} = \min{\{d_\text{move\_block}, d_\text{move\_empty}\}}$. Let $T_{(x, y)} = d_\text{entry}^\text{min} + s_{(x, y)} d_\text{move}^\text{min} + \overline{z}_{(x, y)} d_\text{deliver}^\text{min} + s_{(x, y)} d_\text{move}^\text{min} + d_\text{leave}^\text{min}$ be the duration of the described action sequence for building the column at $(x, y)$ ($\overline{z}_{(x, y)}$ is the desired height of the column). Then $l_r = \max\{T_{(x, y)} \mid \forall (x, y) \in \mathcal{P} \wedge \overline{z}_{(x, y)} > 0\} \leq T_r$ is a MACC lower bound (building of each column is independent in MACC$_r$, due to the relaxations, $l_r$ is equal to the longest duration $T_{(x, y)}$).

A simple mission duration improvement function is proposed to assist in cost-benefit analysis and makespan estimation for more precise action duration mapping of the fraction-time model. Let $d_a^\text{avg} = (\sum_{i \in \mathcal{R}_a}{d_i}) / |\mathcal{R}_a|, a \in \mathcal{A}$, where $|\mathcal{R}_a|$ is the number of elements in set $\mathcal{R}_a$. Let $\alpha = \sum_{a \in \mathcal{A}}{d_a^\text{avg}} / |\mathcal{A}|$ be the estimation of the makespan increase coefficient. Let $T_\text{b}$ be the makespan computed by the fraction-time model where all actions have the duration of one timestep. The coefficient $\alpha$ is proposed to compute the relative makespan increase estimate, when using a more detailed action duration mapping in the fraction-time model. The estimated makespan is defined as $T_h = \max(l_r, \min(u_f, \lceil \alpha T_\text{b} \rceil))$, where $u_f$ is an upper bound defined later in this chapter.

Finally, the fraction-time model can be used for its own makespan estimation. Let $r_i$, $\mathcal{R}$, $T$ and $d_i$ for the fraction-time model with less precise action duration mapping be marked as $r_i'$, $\mathcal{R'}$, $T'$ and $d_i'$, respectively. Since the more precise mapping of action durations makes the model more computationally demanding, a model with less precise mapping -- defined as $\max_{i \in \mathcal{R'}_a}{d_i'} \leq \min_{i \in \mathcal{R}_a}{d_i}, \forall{a \in \mathcal{A}}$ -- can be used for estimating the makespan of the more computationally demanding task. In this regard, the model with $d_i' = 1, \forall i \in \mathcal{R'}$ is especially interesting, as constant action durations should provide the smallest runtime of the fraction-time model for given height-map. Let $l_f = T'$ be a lower bound gained using fraction-time model with less precise action duration mapping.

Let $u_f$ be an upper bound, defined as the execution duration of a plan by the fraction-time model with less precise action duration mapping, where the actions use durations of the more precise mapping and the agents wait at the beginning of each timestep, until all actions that were supposed to end at that timestep are performed. This strategy (padding the plan with wait actions to keep robots synchronized during timestep changes) can also be used when executing the plan on real hardware. When $d_i' = 1, \forall{i \in \mathcal{R'}}$, $u_f$ is defined by equation \ref{eqn:ufsimple}.

\begin{align}
\begin{split}
u_{f_0} = 0\\
u_{f_n} = u_{f_{n-1}} + \\
    \max_{(t_s, t_e, x, \dots, z') \in \mathcal{R'}_{n-1, *, *, *, *, *, *, *, *, *}}{f_d(u_{f_{n-1}}, x, \dots, z')}, \\
\end{split}
\label{eqn:ufsimple}
\end{align}

Let $u_c = T' * \max_{v \in \mathcal{Q}}{f_d(v)}$ be a naive upper bound, where $T'$ is the makespan of fraction-time model with $d_i' = 1, \forall{i \in \mathcal{R'}}$. In practice we multiply the makespan of unit-action-duration fraction-time model with the duration of the longest action.

\section{\uppercase{Model Usage}}
The inspiration for the fraction-time model comes from the TERMES robots and the different mean duration of their high-level instructions. The model generates plans for TERMES-like multi-agent systems through solving the MACC problem. The model is designed to allow the user to describe the action durations at various levels of detail and compute solutions for those durations with optimal makespan and sum-of-costs at given level of detail. The main purpose of this model, however, is to provide a reference solution for future heuristic/sub-optimal models which now can utilize information about different durations of agent action types (in much the same way the original optimal MILP model has been used in \cite{srinivasan2023multiagent} \cite{pddl}). Another potential use-case for the model is creation of reference solutions for models with constraints on agent access and exit position, like \cite{scalableConstructionCompiler}.

The fraction-time model was first proposed as part of a two-part application for multi-agent construction in Minecraft \cite{autoref1}. The model is used to create plans, which are then visualized using Minecraft with agents driven using Malmo API \cite{johnson2016malmo}.

\subsection{Very Long Pick-up Duration}

An interesting concept to study is MACC, where the pick\_up action type has an order of magnitude longer duration, relative to the rest of the action types. The longer disassembly times partially bridge the gap between the methods for solving MACC, which utilize ramps for building the structure, and the ones that do not utilize the ramps. In the real world, the long block detachment times may result from a robot struggling to disconnect a strong inter-block attachment. 

\begin{figure}
    \centering
    \includegraphics[width=0.7\columnwidth]{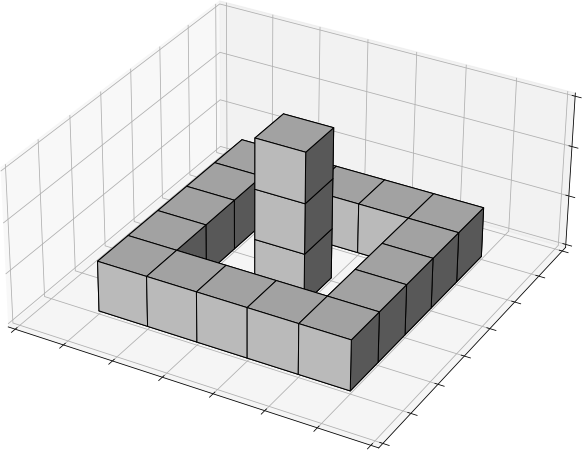}
    \caption{Test instance for long duration block pick-up.}
    \label{fig:longpickup1}
\end{figure}

The figure \ref{fig:longpickup1} shows an instance first used to demonstrate the algorithm in \cite{TACR}. The algorithm heuristically solves the MACC problem in polynomial time by using dynamic programming on a spanning tree, which is optimized in the outer loop. Unlike the fraction-time model, this algorithm aims to heuristically minimize the number of pick\_up and deliver actions \cite{TACR}.

\begin{figure}
    \centering
    \includegraphics[width=0.7\columnwidth]{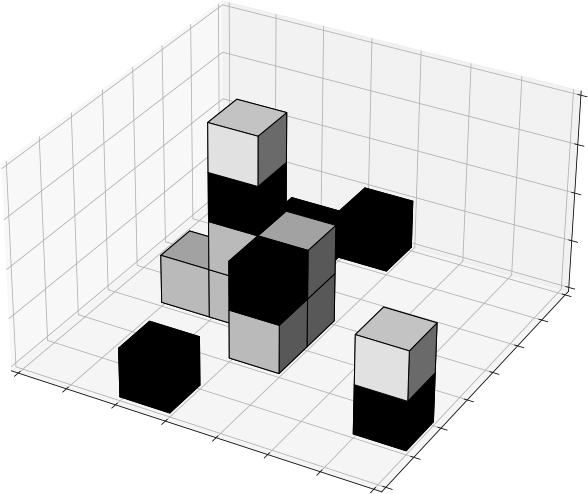}
    \caption{State of the construction at timestep 5; there are two access paths to the middle column, allowing delivery of the third block right after the second block has been placed.}
    \label{fig:longpickup2}
\end{figure}

Because each deliver action, which does not place a block in its final position, must be followed by a pick\_up operation, the optimization problem is equivalent to minimizing the number of pick\_up actions. While the fraction-time model is not built to exactly optimize the count of any action type in particular, by setting $d_\text{pick\_up} >> d_a, \forall a \in \mathcal{A} \setminus \{\text{pick\_up}\}$, the model is forced to limit the number of pick\_up actions to keep the makespan low. Interestingly, this does not produce the minimum number of pick\_up actions possible, as can be demonstrated by solving the toy instance present in the \cite{TACR} with the fraction-time model using $d_\text{pick\_up} = 10$ and $d_a = 1, \forall a \in \mathcal{A} \setminus \{\text{pick\_up}\}$. The resulting plan uses two ramps to reach the center tower at timestep 5, as shown in figure \ref{fig:longpickup2}. The makespan is kept low by parallel deconstruction of the ramps. This behavior can be beneficial when the agents require a long time to detach a block.

\section{\uppercase{Experiments}}
The first experiment aims to demonstrate construction time reduction for non-constant action durations. The experiment is performed with the Gurobi 10.0.3 solver \cite{Gurobi}, limited to 32 threads on an Intel Skylake processor with 16 physical cores and hyper-threading, running at 3 GHz with 132 GB of RAM.

\begin{figure}
    \centering
    \includegraphics[scale=1]{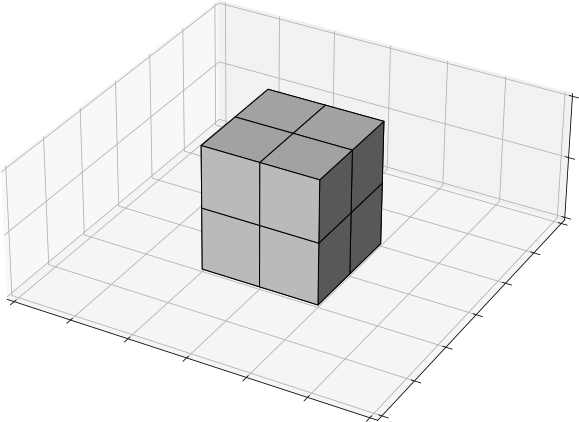}
    \caption{Instance used for makespan estimation precision measurement.}
    \label{fig:approx_instance}
\end{figure}

\begin{figure}
    \centering
    \includegraphics[trim={1.5cm 23cm 2.3cm 1.8cm},clip,width=1.0\columnwidth]{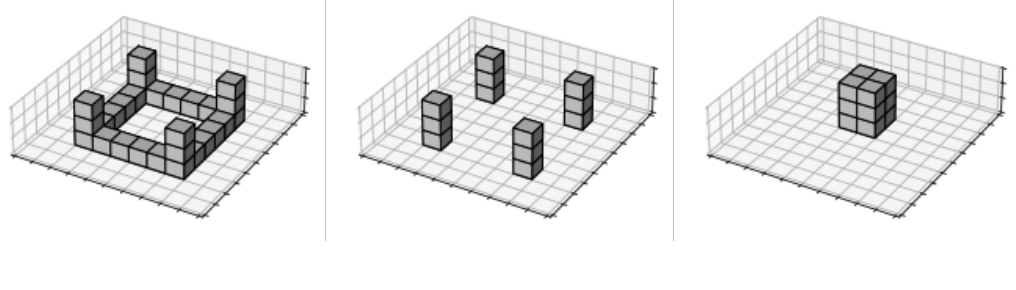}
    \caption{Three different instances used in experiments.}
    \label{fig:instances}
\end{figure}

For evaluation, we translate the fraction-time model to Python Gurobi API, which we use to describe the objective function \ref{eqn:model01} and all the constraints (\ref{eqn:model02} -- \ref{eqn:model16}). The resulting translation requires a known makespan and optimizes only the sum-of-costs. To optimize over makespan as the primary criterion, we add an outer loop, iterating over makespans starting at $l_r$ (lower bound estimate) and increasing by 1 until a solution (construction plan) is found. The input for the Python program is specified as a 2-dimensional height-map of the structure, durations for enter, leave, move\_block, move\_empty, deliver and pick\_up action types. The output of the model is a set of indicators, which action tuples were used. Each action tuple contains start and end position, which is used to assign the actions to agents and compile a list of actions to be performed by each agent. The compilation of this list is done using Python as part of the post processing of the Gurobi solution.

We experiment with three structure instances shown in figure \ref{fig:instances}. These instances represent a subset of partially-modified instances from \cite{koenigexact} that have the lowest solver computation time. The modification consists of removing the flat area without blocks around the third structure to reduce access time for the agents. The main aim of the instance modification is to further reduce computation time because of the expected increased computational load caused by the generalization. The fraction-time model is used to compute plans for construction of each structure instance. Four action duration sets (i.e. sets of action durations) are used for building each structure -- the action durations can be seen in table \ref{tab:experimental-action-durations}. The maximum number of agents is chosen to be 50 for instance 1 and 2; 20 for the third instance (due to smaller construction area).

The table \ref{tab:experimental-results} shows results of the first experiment. The experiment measures solver runtime, solution makespan and sum-of-costs for each instance--action-type-set duo. The construction plans are saved as well, the 1-timestep plan is used for $u_f$ and $u_c$ makespan upper bound computation (result in parentheses next to makespan). The $u_f$ and $u_c$ values are very similar, showing that in most cases, the more easily computed $u_c$ upper bound is sufficient. The makespan values show decrease in construction duration in comparison with wait-action padded plan, according to $u_f$. The construction duration decreases on average by approximately 19\% for {\it 1-2 timestep}, 6\% for {\it 1-2-3 timestep} and 9\% for {\it TERMES timestep}.

Interestingly, the number of agents used in the mission (the \enquote{Robots} column) differs for some tasks. This is likely caused by some sub-plans, which do not belong to the critical path and can be executed in parallel, being executed sequentially in some cases. This can reduce the total number of robots within the building area, as only the maximum number of agents is constrained and more agents are provided than is critical to get the minimum makespan and sum-of-costs.

To confirm that the minimum (leftmost) number of agents in the \enquote{Robots} column is required to achieve the given makespan and sum-of-costs, we rerun the problem instances with the maximum number of agents set to this number and with the maximum number of agents decreased by one. The results match the expectations -- decreasing the number of agents below minimum value causes either an increase in makespan (all action duration sets of instances 1 and 2, 1-timestep and TERMES-timestep action duration sets of instance 3) or sum-of-costs (1-2-timestep and 1-2-3-timestep action duration sets of instance 3).

The table also contains the average value and sampling variance of the runtime, showing a steep growth in computational complexity. The experiment results suggest an exponential increase of computation time in regards to makespan, which was also observed in a similar model by \cite{srinivasan2023multiagent}. This is likely caused by linear dependency between the number of MILP variables ($r_i$ and $h_i$) and model makespan (also noted by \cite{srinivasan2023multiagent}) and a general MILP problem being NP-hard \cite{NPhardMILP}.

\begin{table}[htb]
\centering
\caption{Action type duration for 1-timestep, 1-2-timestep and 1-2-3-timestep action sets.}
\begin{tabular}{lllll}
\hline
&\multicolumn{4}{c}{Action duration sets}
\\ Action & 1 & 1-2 & 1-2-3 & TERMES \\ \hline \hline
\multicolumn{1}{l|}{enter} & 1 & 2 & 3 & 3 \\ \hline
\multicolumn{1}{l|}{leave} & 1 & 1 & 2 & 3 \\ \hline
\multicolumn{1}{l|}{move\_block} & 1 & 1 & 3 & 3 \\ \hline
\multicolumn{1}{l|}{move\_empty} & 1 & 1 & 1 & 2 \\ \hline
\multicolumn{1}{l|}{pick\_up} & 1 & 2 & 3 & 2 \\ \hline
\multicolumn{1}{l|}{deliver} & 1 & 2 & 3 & 3 \\ \hline \hline
\multicolumn{1}{l|}{max} & 1 & 2 & 3 & 3 \\ \hline
\end{tabular}
\label{tab:experimental-action-durations}
\end{table}

\begin{table*}[htb]
\centering
\caption{Experimental results.}
\resizebox{\textwidth}{!}{
\begin{tabular}{llllllll}
\hline
Instance & Action duration set & Run-time & Run-time & Makespan & Makespan & Sum-of-costs & Robots \\
 &  & mean {[}s{]} & sampling & lower & with ($u_f; u_c$ &  &  \\
 &  &  & variance {[}$s^2${]} & bound & upper bound) &  &  \\ \hline \hline
\multirow{3}{*}{1
\begin{minipage}[b]{0.1\textwidth}
    \includegraphics[width=\textwidth]{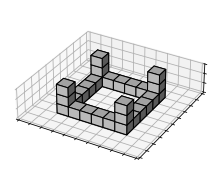}
\end{minipage}
} & 1-timestep & 26.75 & 1.41 & 6 & 11 (11; 11) & 232 & 32, 33, 34, 37 \\
 & 1-2-timestep & 355.33 & 95.16 & 10 & 17 (21; 22) & 316 & 32, 33 \\
 & 1-2-3-timestep & 426.32 & 28.48 & 15 & 30 (32; 33) & 576 & 32 \\
 & TERMES-timestep & 322.73 & 9.19 & 18 & 30 (33; 33) & 648 & 32, 33, 34 \\ \hline
\multirow{3}{*}{2
\begin{minipage}[b]{0.1\textwidth}
    \includegraphics[width=\textwidth]{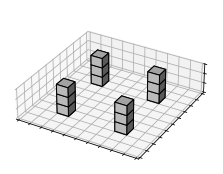}
\end{minipage}
} & 1-timestep & 24.10 & 1.05 & 6 & 11 (11; 11) & 196 & 32 \\
 & 1-2-timestep & 58.88 & 0.93 & 10 & 17 (21; 22) & 284 & 32 \\
 & 1-2-3-timestep & 287.24 & 15.76 & 15 & 30 (32; 33) & 508 & 32 \\
 & TERMES-timestep & 264.28 & 2.83 & 18 & 30 (33; 33) & 548 & 32 \\ \hline
\multirow{3}{*}{3
\begin{minipage}[b]{0.1\textwidth}
    \includegraphics[width=\textwidth]{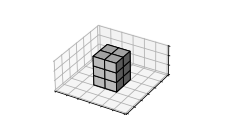}
\end{minipage}
} & 1-timestep & 239.21 & 47.98 & 6 & 14 (14; 14) & 142 & 16, 17 \\
 & 1-2-timestep & 303.92 & 43.62 & 10 & 21 (27; 28) & 213 & 17 \\
 & 1-2-3-timestep & 3807.62 & 414.87 & 15 & 37 (41; 42) & 352 & 17 \\
 & TERMES-timestep & 6152.27 & 1132.99 & 18 & 38 (41; 42) & 398 & 16, 17 \\ \hline
\end{tabular}
}
\label{tab:experimental-results}
\end{table*}

The second experiment aims to evaluate the makespan estimation function $T_h$, namely the effects of varying action durations on makespan estimation accuracy. For this purpose, all action type durations, without the $T_\text{wait}$, are assigned durations $T_a \in \{1, 2, 3\} = \beta, \forall a \in \mathcal{A} \setminus \{\text{wait}\}$ and all combinations of those durations are measured (requiring $|\beta|^{|A|-1} = 729$ construction plans to be computed). $T_\text{wait}$ is left as 1, because the minimum time a robot can wait is in general not limited. Only one target structure is used -- a $2\times2\times2$ cube (see figure \ref{fig:approx_instance}). This very small structure is selected due to the high number of required measurements.

\begin{figure}
    \centering
    \includegraphics[width=\columnwidth]{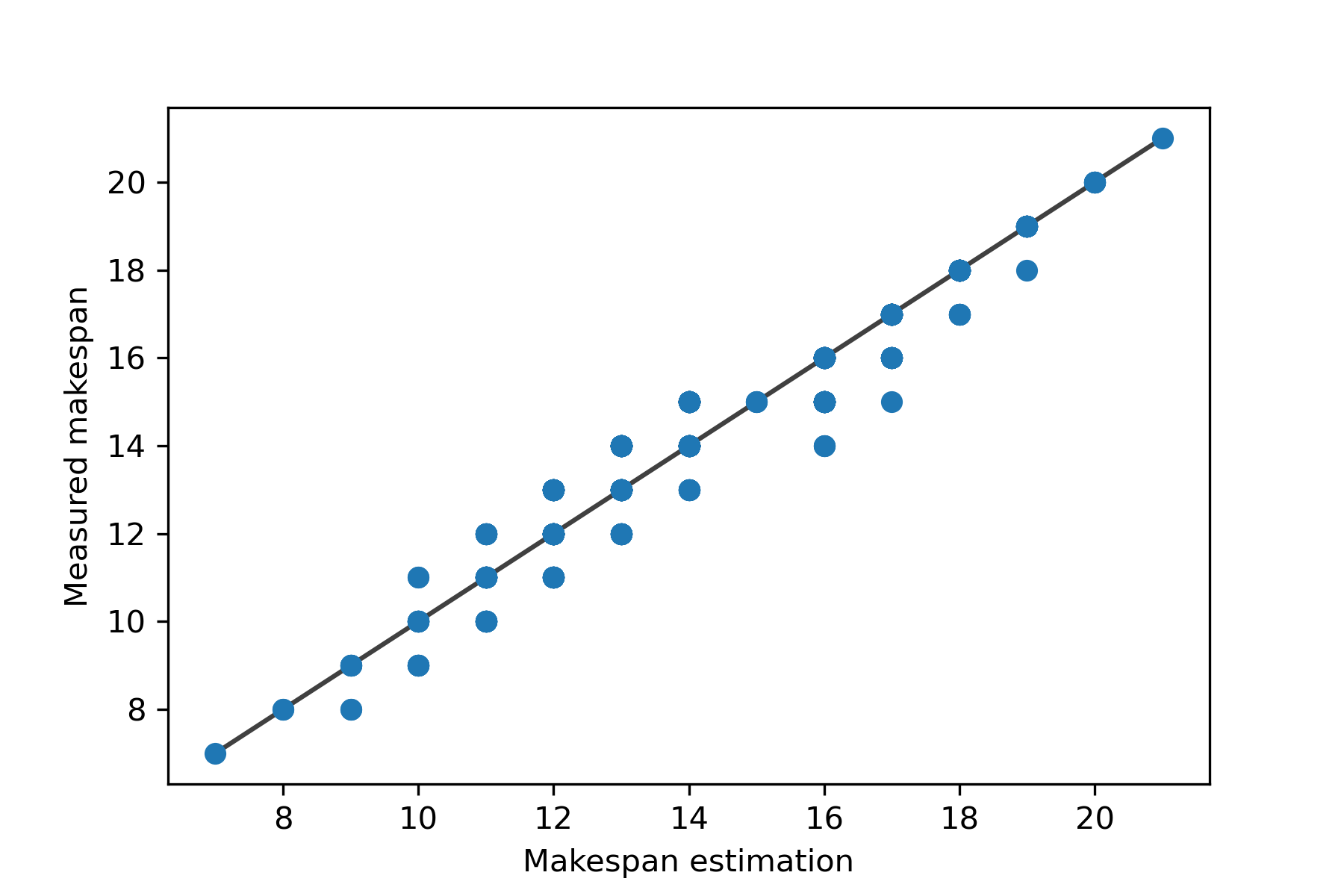}
    \caption{Scatter plot of makespan\-heuristic relation.}
    \label{fig:makespanheuristic}
\end{figure}

The experiment results in figure \ref{fig:makespanheuristic} show that the makespan estimation both can under- and over-estimate the actual makespan value. The root mean square error (RMSE) of the makespan estimation function is 0.655, with minimum estimation error absolute value, relative to makespan, being 0 and maximum 0.143. The error stems from simplicity of the makespan estimation function -- it depends only on action type duration mean and makespan of the generalized model with unit action durations. It does not take into account the usage of the action types.

For instance, \enquote{deliver} type action is used mainly for building the target structure and ramps to access its higher levels. Pick-up action is mainly used to disassemble the ramps. In the earlier work \cite{koenig2017case}, it is noted that both actions combined can be used to create a bucket-brigade-like behavior, where agents form a line with one block wide free spaces between agents. By repeatedly picking up blocks and delivering them to the other side of the robot, the robots can transport blocks while limiting the number of move actions.

Move\_block action is mainly used for transportation of blocks belonging to both structure and ramps from robot start position to their intended placement location and for transportation of ramp blocks out of the building area. Move\_empty action is for leaving the building area, when the agent is not assigned any ramp block to move out of the building area.

Enter action lets the agents enter the building area from its border cells. Agents can enter with blocks, serving as suppliers of the construction material. Leave action lets the agents leave the building area when standing atop a border cell. Leaving frees up the agent to enter at any border cell, because the enter action can be performed only if the entering agent would not violate the maximum number of agents constraint. They can leave while carrying a block (used for deconstruction of the ramps). Enter and leave actions always form a pair, mainly due to constraints specifying that there are no agents within the building area at the start and end of the construction plan, so all agents who enter, must also leave.

\subsection{Agent Movement Speed Dependent on Vertical Position}

The last test aims to demonstrate another beneficial action duration assignment for the fraction-time model. The real TERMES agents use marine foam blocks with indentations/protrusions and neodymium magnets to ensure block alignment and stability \cite{petersen2011termes}. The resulting block columns are stable enough that the agents are not required to adapt their movement speed to the height of the structure they are traversing -- at least at the scales of structures used in the paper. For cases of TERMES-like multiagent systems with less stability, the fraction-time model allows the adjustment of the agent movement speed according to the vertical position of the agent. This is demonstrated on a task with action duration function $f_d$ defined by the equation \ref{eqn:h_v_fd}. The equation ensures that the robot action durations grow linearly with the vertical position of the robot at the end of the action (with the exception of the wait action type, where the robot stays still).

The experiment measures 10 run-times with action durations $f_{d_h}$ (referred to as \enquote{Unstable columns} in the results) on the instance 1. Makespan, sum-of-costs and run-time are compared with results for \enquote{TERMES-timestep} action duration set of instance 1 (see table \ref{tab:experimental-results}. The \enquote{TERMES-timestep} results are referred to as \enquote{Base} in the table \ref{tab:h_v_dependency_task}). The table \ref{tab:h_v_dependency_task} shows the results of the experiment. The makespan of unstable columns task has increased by 40\%, the sum-of-costs by approximately 7.4\%, relative to the base. The increase of makespan was expected, due to the $f_{d_h}$ consisting of TERMES-timestep $f_d$ with an added positive component, linearly dependent on agent vertical position. The interesting part of the result is the small 7.4\% relative increase in sum-of-costs in relation to the 40\% relative increase in makespan. This indicates a lower agent utilization in case of the unstable columns task.

The run-time increased with the makespan. However, it is still notably lower than runtime of instance 3 with TERMES-timestep in table \ref{tab:experimental-results}, which has makespan 38. This indicates, that while run-time is greatly dependent on makespan, the remaining task-dependent constraints, such as the structure height map and the agent count, also affect the computational complexity to a relatively large degree.

\begin{multline}
    f_{d_h}(v) =
     \begin{cases}
       3 &\quad\text{if } v \in \mathcal{Q}_\text{entry} \cup \mathcal{Q}_\text{leave}\\
       2+z' &\quad\text{if } v = (\dots, z') \in \mathcal{Q}_\text{move\_empty}\\
       3+z' &\quad\text{if } v = (\dots, z') \in \mathcal{Q}_\text{move\_block}\\
       2+2z' &\quad\text{if } v = (\dots, z') \in \mathcal{Q}_\text{pick\_up}\\
       3+2z' &\quad\text{if } v = (\dots, z') \in \mathcal{Q}_\text{deliver}\\
       1 &\quad\text{otherwise.}
     \end{cases}\\
     \label{eqn:h_v_fd}
\end{multline}

\begin{table}
    \centering
    \caption{Comparison of TERMES task solution where movement and block manipulation speed depends on agent height (unstable columns) with task solution, where movement speed is independent of the agent height (base).}
    \begin{tabular}{ccc} \hline
        TERMES & Base & Unstable columns \\ \hline
        Makespan & 30 & 42 \\
        Sum-of-costs & 648 & 696 \\
        Mean run-time [s] & 322.73 & 3837.64 \\ \hline
    \end{tabular}
    \label{tab:h_v_dependency_task}
\end{table}

\section{\uppercase{Conclusion}}
A new branch of the multi-agent construction has recently emerged -- exact optimization of the problem. The current state-of-the-art exact approach provides solutions with optimal makespan and sum-of-costs but assumes all agent high-level actions to have the same duration. This is not the case in the real world. For instance, the mean duration of block manipulation actions of the TERMES robot differs by 46\%.

This has motivated us to generalize the current state-of-the-art exact MILP model \cite{koenigexact} into the fraction-time model with action durations specified as part of the model input. The generalization is non-trivial -- upholding the agent collision constraint has necessitated the redefinition of agent collision using exclusion zones while keeping the model acceptably computationally demanding required the preservation of the two network flow structures within the model. We test the generalized model on various assigned action durations (including TERMES action durations), studying its behavior. The experiment shows a 9\% decrease in construction duration (makespan vs $u_f$ upper bound, which can provide a non-optimal plan)  for the TERMES robots. Other action duration sets also show similar results.

The second experiment analyses the heuristic makespan estimation function $T_h$ and its behavior with different action duration sets. The experiment used toy instance -- $2 \times 2 \times 2$ and showcased the spread of estimated makespan values in comparison with the real ones. The RMSE of the $T_h$ is 0.665, making it a viable source of information for cost-benefit analysis when considering a more computationally demanding action duration mapping.

Two experiments also demonstrated the flexibility of the fraction-time model -- in the first one, an assignment was given to the model, where the pick\_up action was an order-of-magnitude slower to execute than other actions. The experiment results indicate that the model utilizes multiple agents for parallel ramp construction and deconstruction to minimize sequential pick\_up actions while minimizing the overall makespan. The last experiment showcased the ability of the model to adjust to agents moving slower at greater heights. The optimal solutions show a notably lower relative utilization of the agents in comparison with the assignment, where agents move at the same speed across all heights.

While the model is computationally demanding, it can provide reference solutions for both existing and future heuristic models. These solutions can be used to estimate the heuristic model efficiency and as long-term targets for models optimizing makespan (i.e. construction duration).

\section*{\uppercase{Acknowledgements}}
This work has been supported by the project number 22-31346S of the Czech Science Foundation GA ČR and by the CTU project SGS23/210/OHK3/3T/18.

\bibliographystyle{apalike}
{\small
\bibliography{bibliography}}

\begin{thebibliography}{}

\bibitem[Barros~dos Santos et~al., 2018]{Quadrotors}
Barros~dos Santos, S.~R., Givigi, S., Nascimento, C.~L., Fernandes, J.~M.,
  Buonocore, L., and de~Almeida~Neto, A. (2018).
\newblock Iterative decentralized planning for collective construction tasks
  with quadrotors.
\newblock {\em Journal of Intelligent {\&} Robotic Systems}, 90(1):217--234.

\bibitem[Bulut and Ralphs, 2021]{NPhardMILP}
Bulut, A. and Ralphs, T.~K. (2021).
\newblock On the complexity of inverse mixed integer linear optimization.
\newblock {\em {SIAM} Journal on Optimization}, 31(4):3014--3043.

\bibitem[Cai et~al., 2016]{multiagentTACR}
Cai, T., Zhang, D.~Y., Kumar, T.~S., Koenig, S., and Ayanian, N. (2016).
\newblock Local search on trees and a framework for automated construction
  using multiple identical robots.
\newblock In {\em Proceedings of the 2016 International Conference on
  Autonomous Agents \& Multiagent Systems}, pages 1301--1302.

\bibitem[Deng et~al., 2019]{scalableConstructionCompiler}
Deng, Y., Hua, Y., Napp, N., and Petersen, K. (2019).
\newblock A compiler for scalable construction by the termes robot collective.
\newblock {\em Robotics and Autonomous Systems}, 121:103240.

\bibitem[{Forgemind ArchiMedia}, 2014]{ImageTERMES}
{Forgemind ArchiMedia} (2014).
\newblock Termes robot 01.

\bibitem[{Gurobi Optimization, LLC}, 2023]{Gurobi}
{Gurobi Optimization, LLC} (2023).
\newblock {Gurobi Optimizer Reference Manual}.

\bibitem[Jenett and Cheung, 2017]{NASAinchworm}
Jenett, B. and Cheung, K. (2017).
\newblock Bill-e: Robotic platform for locomotion and manipulation of
  lightweight space structures.
\newblock In {\em 25th AIAA/AHS Adaptive Structures Conference}, page 1876.

\bibitem[Johnson et~al., 2016]{johnson2016malmo}
Johnson, M., Hofmann, K., Hutton, T., and Bignell, D. (2016).
\newblock The malmo platform for artificial intelligence experimentation.
\newblock In {\em IJCAI}, pages 4246--4247. Citeseer.

\bibitem[Koenig and Kumar, 2017]{koenig2017case}
Koenig, S. and Kumar, S. (2017).
\newblock A case for collaborative construction as testbed for cooperative
  multi-agent planning.
\newblock In {\em Proceedings of the ICAPS-17 Scheduling and Planning
  Applications Workshop (SPARK)}.

\bibitem[Kumar et~al., 2014]{TACR}
Kumar, T.~K., Jung, S., and Koenig, S. (2014).
\newblock A tree-based algorithm for construction robots.
\newblock {\em Proceedings of the International Conference on Automated
  Planning and Scheduling}, 24(1):481--489.

\bibitem[Lam et~al., 2020]{koenigexact}
Lam, E., Stuckey, P., Koenig, S., and Kumar, T. (2020).
\newblock Exact approaches to the multi-agent collective construction problem.
\newblock In Simonis, H., editor, {\em Principles and Practice of Constraint
  Programming}, Lecture Notes in Computer Science, pages 743--758. Springer.
\newblock International Conference on Principles and Practice of Constraint
  Programming 2020, CP2020 ; Conference date: 07-09-2020 Through 11-09-2020.

\bibitem[Petersen et~al., 2011]{petersen2011termes}
Petersen, K.~H., Nagpal, R., and Werfel, J.~K. (2011).
\newblock Termes: An autonomous robotic system for three-dimensional collective
  construction.
\newblock {\em Robotics: science and systems VII}.

\bibitem[Pinciroli et~al., 2020]{studentInchworm}
Pinciroli, C., Lewin, G., Cowlagi, R., Huang, X., Collins, C., Contreras, J.,
  Dhanaraj, N., Liang, H., Rizzo, T., and Wagner, C. (2020).
\newblock Swarm construction: A method in multi-agent robotic assembly.
\newblock {\em Worcester Polytechnic Institute}.

\bibitem[Piranda and Bourgeois, 2018]{SphereRobots}
Piranda, B. and Bourgeois, J. (2018).
\newblock {\em Geometrical Study of a Quasi-spherical Module for Building
  Programmable Matter}, pages 387--400.
\newblock Springer International Publishing, Cham.

\bibitem[Rameš, 2021]{autoref1}
Rameš, M. (2021).
\newblock Compilation of multi-agent collective construction in the minecraft
  game.
\newblock Bachelor thesis, Czech Technical University in Prague, Faculty of
  Information Technology, Czech Republic.

\bibitem[Romanishin et~al., 2013]{CubeRobots}
Romanishin, J.~W., Gilpin, K., and Rus, D. (2013).
\newblock M-blocks: Momentum-driven, magnetic modular robots.
\newblock In {\em 2013 IEEE/RSJ International Conference on Intelligent Robots
  and Systems}, pages 4288--4295.

\bibitem[Singh et~al., 2023]{pddl}
Singh, S., Gutow, G., Srinivasan, A.~K., Vundurthy, B., and Choset, H. (2023).
\newblock Hierarchical propositional logic planning for multi-agent collective
  construction.
\newblock In {\em Construction Robotics Workshop}.

\bibitem[Srinivasan et~al., 2023]{srinivasan2023multiagent}
Srinivasan, A.~K., Singh, S., Gutow, G., Choset, H., and Vundurthy, B. (2023).
\newblock Multi-agent collective construction using 3d decomposition.

\end{thebibliography}
\end{document}